\documentclass[12pt]{iopart}

\usepackage[pdftex]{graphicx}
\usepackage{dcolumn}
\usepackage[latin1]{inputenc}
\usepackage{iopams}
\usepackage{color}
\usepackage{textcomp}
\usepackage{iopams}

\newcommand{\ud}[1]{{#1^{\dagger}}}

\newcommand{\mean}[1]{\langle#1\rangle}

\newcommand{\nn}{\nonumber}

\begin{document}


\title{Correlator expansion approach to stationary states of weakly
  coupled cavity arrays}

\author{Elena del Valle and Michael J. Hartmann}
\address{Physik Department, Technische
  Universit\"at M\"unchen, James-Franck-Stra{\ss}e, 85748 Garching,
  Germany}

\date{\today}

\begin{abstract}
  We introduce a method for calculating the stationary state of a
  translation invariant array of weakly coupled cavities in the
  presence of dissipation and coherent as well as incoherent drives.
  Instead of computing the full density matrix our method directly
  calculates the correlation functions which are relevant for
  obtaining all local quantities of interest. It considers an
  expansion of the correlation functions and their equations of motion
  in powers of the photon tunneling rate between adjacent cavities,
  leading to an exact second order solution for any number of
  cavities. Our method provides a controllable approximation for weak
  tunneling rates applicable to the strongly correlated regime that is
  dominated by nonlinearities in the cavities and thus of high
  interest.
\end{abstract}

\maketitle

\section{Introduction}

The study of light matter interactions that are enhanced by confining
light fields in electromagnetic cavities has been a thriving
discipline of Quantum Optics throughout the last decades. Particularly
with the advent of realizations of the so called strong coupling
regime where the strength of the interaction between a photon and a
quantum emitter exceeds the decay mechanisms for photons and emitters,
experimental investigations of the coherent interactions between
single emitters and individual photons became
possible~\cite{raimond2001}.
 
In recent years, a new direction of cavity quantum electrodynamics
(cavity-QED) has developed, in which multiple cavities that are
coupled via the exchange of photons are considered. Such setups are
particularly intriguing if the cavities are connected forming an array
and the strong coupling regime is achieved in each cavity of the
array. These devices would then give rise to quantum many-body systems
of strongly interacting photons and
polaritons~\cite{hartmann2006,angelakis2007,greentree2006}. As an
alternative to a cavity array, one may also consider optical fibers
that couple to nearby atoms~\cite{CGM+08,KH09} or even clouds of
Rydberg atoms that are optically thick in free space
\cite{Peyronel2012}. Both these systems avoid the need to build
mutually resonant cavities, which is possible \cite{Ritter2012} but
can be rather challenging in the optical range.  For microwave photons
it is however perfectly feasible to build large arrays of mutually
resonant cavities on one chip in an architecture known as
circuit-QED~\cite{houck2012,Lucero2012}.

For strongly interacting polaritons and photons in coupled arrays of
micro-cavities and optical fibers, possibilities to observe
equilibrium phenomena, such as a Mott
insulator~\cite{hartmann2006,angelakis2007,greentree2006} or a
Tonks-Girardeau gas~\cite{CGM+08,KH09}, have mostly been addressed so
far and the development has been summarized in the
reviews~\cite{hartmann2008,tomadin2010,houck2012}.  In every
experiment that involves light-matter interactions, some photons will
however inevitably be lost from the structure due to imperfect light
confinement or emitter relaxation so that thermal states are no longer
an appropriate description of the system.  To compensate for such
losses, coupled cavity arrays are thus most naturally studied in a
regime where a coherent or incoherent input continuously replaces the
dissipated excitations. This mode of operation eventually gives rise
to a driven dissipative regime, where the dynamical balance of loading
and loss processes leads to the emergence of stationary states. Yet
the properties of stationary states of driven dissipative systems are
only explored to a much lesser degree than the properties of thermal
equilibrium states. For coupled cavities, small arrays have been
considered exactly \cite{carusotto2009} and mean field approaches for
larger arrays have been employed \cite{nissen2012}. Moreover numerical
studies found signatures for crystallization \cite{Hartmann10} and
photon solids were predicted for arrays with cross Kerr interactions
\cite{Jin13}.

As driven dissipative quantum many-body systems have to date only
barely been explored, there is a need for technical tools for their
efficient description.  Here we introduce a perturbative technique for
the calculation of the physical properties of stationary states of
driven dissipative cavity arrays.  Our approach assumes a large,
translation invariant cavity array where the quantum states of all
cavities are identical. Instead of computing the full density matrix,
it directly calculates the correlation functions which are relevant
for obtaining all local quantities of interest. A power expansion of
the formal exact solution to second order in the photon tunneling rate
between adjacent cavities, provides semi-analytical results for any
number of cavities.

\section{Exact solution in the steady state}

We consider an array of cavities that are coupled via mutual photon
tunneling and where each cavity is doped with a Kerr nonlinear medium
that generates a strong photon-photon interactions. In a frame that
rotates at the frequency of the coherent drive lasers, this system is
described by a Bose-Hubbard Hamiltonian with local coherent drives
($\hbar=1$),
\begin{eqnarray}
  \label{eq:ThuJan10202154CET2013}
  H_N&=&\sum_{i=1}^N \Big[\Delta a^\dagger_i a_i+\frac{U}{2}a^\dagger_i
  a^\dagger_i a_i a_i +\Omega( a^\dagger_i +a_i)\Big] \nn \\
  &+&J \sum_{i=1}^{N-1}  (a^\dagger_i a_{i+1}+a^\dagger_{i+1} a_i) + J (a^\dagger_N a_1+a^\dagger_1 a_N) 
\end{eqnarray}
where we assumed periodic boundary conditions so that $N$ cavities
form a circle coupling to nearest left and right neighbors. Here,
$\Delta = \omega_a - \omega_\mathrm{L}$ is the detuning of the cavity
resonance frequency $\omega_a$ from the laser frequency
$\omega_\mathrm{L}$, $U$ is the strength of the Kerr nonlinearity, $J$
the rate of photon tunneling between the cavities and $\Omega$ the
drive amplitude of a coherent laser drive. We consider the local
energy scales orders of magnitude smaller than the cavity frequency,
that is, $U$, $J$, $\Delta$, $\Omega \ll \omega_a$, so that the
rotating wave approximation can be applied. Adding decay and
incoherent pumping of the modes, the total Liouvillian that describes
the dynamics of this system reads
\begin{eqnarray}
  \label{eq:ThuJan10202150CET2013}
  \partial_t \tilde\rho  =i[\tilde\rho,H_N]  &+&\frac{\gamma}2\sum_{i=1}^{N} (2a_i
  \tilde\rho\ud{a_i}-\ud{a_i} a_i \tilde\rho-\tilde\rho\ud{a_i} a_i) \nonumber\\
  &+&\frac{P}2\sum_{i=1}^{N} (2\ud{a_i} \tilde\rho a_i-a_i\ud{a_i}
  \tilde\rho-\tilde\rho a_i\ud{a_i})\,,
\end{eqnarray}
where $\tilde \rho$ denotes the total density matrix of the cavity
array. $\gamma$ is the rate of photon decay and $P$ the rate at which
photons are incoherently pumped into the device. We are interested in
the steady state under a continuous excitation of each of the
cavities, represented by the reduced density matrix $\rho$ of a single cavity.
As all cavities have exactly the same properties and dynamics, we find
$\rho_{1}=\rho_2=\ldots=\rho_N=\rho$, where $\rho_{j}$ is the reduced
density matrix of cavity number $j$.  

Here, instead of obtaining the
full density matrix of the array in the steady state
($\partial_t\tilde \rho=0$) and then tracing out all cavities but one,
to obtain $\rho$, we compute the steady state properties directly in
the form of the local mean values of all possible operators defining the
system. These can be written as
\begin{equation}
  \label{eq:SatMar9150643CET2013}
\mean{a_i^{\dagger m}a_i^{n}}=\Tr (\tilde\rho a_i^{\dagger m}a_i^{n}) 
\end{equation}
where $m$ and $n$ are integers.  Let us call~$\mathcal{O}$ the set of
operators the averages of which correspond to the correlators required
to describe the full $N$-cavity system, i.e., $\mathcal{O}$ includes
all the sought observables as well as operators which couple to them
through the equations of motion, $\mean{a_1^{\dagger
    m}a_1^{n}a_2^{\dagger \mu} a_2^{\nu}\ldots a_N^{\dagger\alpha}
  a_N^{\beta}}$. To simplify notation, we express this general
correlator as $\{\{m,n\},\{\mu,\nu\}\ldots
\{\alpha,\beta\}\}$. Naturally, in the case of an anharmonic mode or
in the presence of any anharmonicity, one must choose an appropriate
truncation in the number of excitations, i.e. $n,m \le
n_\mathrm{max}$, that also truncates the number of correlations in
this set. For a given driving intensity, the truncation must be high 
enough to yield converged, accurate results.

From the master equation~(\ref{eq:ThuJan10202150CET2013}) one can
obtain the set of coupled equations for the full set of
operators~$\mathcal{O}$, which we
write in the matrix form,
\begin{equation} 
  \label{eq:matrixform}
  \partial_t \tilde v = \tilde M \tilde v+\tilde I\,,
\end{equation}
with all correlators forming the vector $\tilde v$,
i.e. $\tilde{v}^{\textrm{T}} = (\mean{a_1}, \mean{a^\dagger_1}, \dots,
\mean{a_1 a_2}, \mean{a^\dagger_1 a_2^2 } \dots )$, 
where the exponent T denotes the transpose. The coefficient
matrix $\tilde M$ and vector $\tilde I$ are derived from the master
equation in a systematic way~\cite{delvalle_book10a,delvalle12a}, as
we explicitly show in the Appendix.  The solution of
Eq.~(\ref{eq:matrixform}) is completely equivalent to computing the
correlators as in Eq.~(\ref{eq:SatMar9150643CET2013}), from the
density matrix obtained by solving
Eq.~(\ref{eq:ThuJan10202150CET2013}).  The exact solution in the
steady state (if it is unique and exists) simply reads
\begin{equation} \label{eq:exactsol}
\tilde v= -\tilde{M}^{-1} \tilde I
\end{equation}

We can considerably reduce the number of operators to a minimal set by
making use of the translational symmetry in the 1D chain
(circle). That is, all correlators that are left or right circular
rotations of the elements in $\{\{m,n\},\{\mu,\nu\}\ldots
\{\alpha,\beta\}\}$, such as
$\{\{\alpha,\beta\},\{m,n\},\{\mu,\nu\}\ldots \}$ or $\{
\{\alpha,\beta\},\ldots,\{\mu,\nu\},\{m,n\}\}$, are redundant because
they are exactly the same. There is, therefore, a maximum of $2N$
representations of the same correlator (less if some of the pairs are
$\{0,0\}$ or mutually equal). We can choose an arbitrary rule to
systematically keep only one representative of such set of redundant
correlators, for instance, we choose the ones where:
\begin{enumerate}
\item The nonzero sets are always to the left and as cluttered
  together as possible, such as $\{\{4,2\},\{3,0\},\{0,0\}\ldots\}$
\item The largest sum of indexes is most to the left: $m+n\geq
  \mu+\nu \geq \ldots \geq \alpha+\beta$, such as $\{ \{3,3\},\{
  1,3\},\{ 0,2\}, \{1,0\}\}$. Together with the previous rule, this may
  give things like $\{ \{ 1,3\},\{3,3\},\{1,0\},\{ 0,0\},\ldots \}$. 
\item If there are two pairs with an equal sum, the one with the
  largest first index is left-most $m\geq \mu$. Together with the
  previous rule, this may give things like $\{ \{
  4,2\},\{3,3\},\{ 0,0\},\ldots \}$.
\end{enumerate}
With this, the vectorial Hilbert space is significantly reduced, for
instance from 6559 to 1033 for $N=4$ and $n_\mathrm{max}=2$. From
$\tilde{v}$ we thus extract a new vector $v$ that only contains the
minimal set of correlators. The equation of motion for $v$ is then
written in terms of a new matrix, $M$, reconstructed by removing the
redundant rows and summing the coefficients of the redundant columns
of $\tilde{M}$, and a new vector, $I$ removing the redundant rows of
$\tilde I$. The exact stationary state solution of the reduced system
of equations is now given by
\begin{equation}
  \label{eq:ThuJan10223943CET2013}
  v= - M^{-1} I\,.
\end{equation}

\subsection{One cavity, $N=1$, and the uncoupled limit}

In the uncoupled limit, $J = 0$, the calculation can be reduced to a
single cavity.  The operators for each cavity, $\mean{a_i^{\dagger
    m}a_i^{n}}$, are the same for all $i=1,\ldots,N$, so let us denote
them by a common name, $\mean{a^{\dagger m}a^{n}}=\{\{m,n\}\}$,
without any index. We can define the vector $v_a$ of all possible
individual cavity operators,
\begin{equation}
  \label{eq:ThuJan10202139CET2013}
  v_a^\mathrm{T} = \left(\mean{a},\mean{\ud{a}},\mean{\ud{a}a}, \dots \right)\,.
\end{equation}
In the case of one cavity or many which are uncoupled, the full
ensemble vector $v$ reduces trivially to $v=v_a$. The equation of
motion for this system reads $\partial_t v_a = M_a v_a+I_a$ and its
stationary solution is,
\begin{equation}
  \label{eq:FriMar1184432CET2013}
  v_a=- M_a^{-1} I_a\,.
\end{equation}

\subsection{$N\leq 4$ cavities}

Four is the minimum number of identical cavities needed to obtain a
general solution to second order in $J$ that is valid for any $N$. The
reason is that four coupled systems imply a qualitative change, as
compared to two or three, as each of them is no longer in contact with all
the others. This represents the general case to second order in $J$.

In this case, the vector of correlators $v$ not only contains the subset $v_a$ with
$\{\{m,n\},\{0,0\},\{0,0\},\{0,0\}\}$, but also another four new
subsets that include cross correlations with two, three or four
cavities. The first one, which we call $v_b$, includes correlators
with two cavities, $\mean{a_1^{\dagger m}a_1^{n}a_2^{\dagger
    \mu}a_2^{\nu}}=\mean{a_2^{\dagger m}a_2^{n}a_1^{\dagger
    \mu}a_1^{\nu}}=\ldots=\mean{a^{\dagger m}a^{n}b^{\dagger \mu}b^{\nu}}=\{
\{m,n\},\{\mu,\nu\},\{0,0\},\{0,0\}\}$. We denote with $b$ the photon
annihilation operator for the second
cavity, and apply the rules to extract the minimal set of operators
($m+n\geq \mu+\nu$, with $m\geq \mu$ in case of degeneracy of the
sum).  For example, for a truncation in each cavity system with
$n_\mathrm{max}=2$ photons, we thus have a dimension of 8 for $v_a$
and 36 for $v_b$.  A third subset, $v_c$, includes correlators with
three cavities, $\mean{a_1^{\dagger m}a_1^{n}a_2^{\dagger
    \mu}a_2^{\nu}a_3^{\dagger p}a_3^{q}}=\mean{a_2^{\dagger
    m}a_2^{n}a_1^{\dagger \mu}a_1^{\nu}a_3^{\dagger
    p}a_3^{q}}=\ldots=\mean{a^{\dagger m}a^{n}b^{\dagger
    \mu}b^{\nu}c^{\dagger p}c^{q}}=\{
\{m,n\},\{\mu,\nu\},\{p,q\},\{0,0\}\}$. Similarly to $b$, we denote with $c$
the photon annihilation operator in the third cavity, where we have applied the circular rules to
obtain the minimal set of operators. Finally, specifically to $N=4$,
we need to consider $v_d$, which includes correlators with 4 cavities
$\mean{a_1^{\dagger m}a_1^{n}a_2^{\dagger \mu}a_2^{\nu}a_3^{\dagger
    p}a_3^{q} a_4^{\dagger s}a_4^{t}}=\ldots=\mean{a^{\dagger
    m}a^{n}b^{\dagger \mu}b^{\nu}c^{\dagger p}c^{q} d^{\dagger
    s}d^{t}}=\{ \{m,n\},\{\mu,\nu\},\{p,q\},\{t,s\}\}$, 
$d$ is the photon annihilation operator in the fourth cavity, and $v_e$
which includes operators of two cavities at alternate positions,
$\mean{a_1^{\dagger m}a_1^{n}a_3^{\dagger
    p}a_3^{q}}=\mean{a_2^{\dagger m}a_2^{n}a_4^{\dagger
    p}a_4^{q}}=\ldots=\mean{a^{\dagger m}a^{n}c^{\dagger p}c^{q}}=\{
\{m,n\},\{0,0\},\{p,q\},\{0,0\}\}$.

We can rewrite Eq.~(\ref{eq:matrixform}) in terms of these subsets of
correlators, each with a different dimension, as five coupled matrix 
equations,
\begin{eqnarray}
  \label{eq:ThuJan10221804CET2013}
  \partial_t v_a &=&(M_a +iJ S_a) v_a +I_a +iJ R_{ab}v_b\,,\label{eq:SunMar10123437CET2013}\\
  \partial_t v_b &=&(M_b +iJ S_b) v_b + (B_{ba} +iJ R_{ba}) v_a +iJ R_{bc}v_c\,, \label{eq:SunMar10123502CET2013}\\
  \partial_t v_c &=&(M_c +iJ S_c) v_c + (B_{cb} +iJ R_{cb})v_b +iJ R_{cd}v_d+(B_{ce}+iJR_{ce})v_e\,, \\
  \partial_t v_d &=&(M_d +iJ S_d) v_d + (B_{dc} +iJ R_{dc})v_c\,, \\
  \partial_t v_e &=&(M_e +iJ S_e) v_e + (B_{ea} +iJ R_{ea})v_a+iJR_{ec}v_c\,.\label{eq:TueMar12191621CET2013}
\end{eqnarray}
Here, we have separated the effect of the hopping $J$ into the
self-renormalization matrices $S$, and the linking matrices $R$, that
only contain integer numbers. The vector $I_a$ and matrices $B$,
contain only the driving parameters $\Omega$ and $P$ (coherent or
incoherent). Other internal parameters such as $\Delta$, $\gamma$ and
$U$ enter in the matrices $M$.

One can solve these equations recurrently in the steady state, from
bottom to top. This may be useful for a small number of cavities where
the expressions are simple. For instance, for $N=2$, the system
reduces to $v_a$ and $v_b$ with
Eqs.~(\ref{eq:SunMar10123437CET2013})--(\ref{eq:SunMar10123502CET2013}),
and solutions:
\begin{eqnarray}
  \label{eq:ThuJan10205937CET2013}
  v_a &=&- \left(M_a +iJ S_a+iJ R_{ab}F_{ba}\right)^{-1} I_a\,,\\
  v_b &=& F_{ba}v_a \,,
\end{eqnarray}
where $F_{ba}=-\left(M_b +iJ S_b\right)^{-1} (B_{ba} +iJ R_{ba})v_a
$. Similarly, for $N=3$ we have:
\begin{eqnarray} 
  \label{eq:ThuJan10214536CET2013}
  v_a &=&- \left(M_a +iJ S_a+iJ R_{ab}F_{ba}\right)^{-1} I_a\,,\\
  v_b &=&F_{ba}v_a \,,\quad\mathrm{and}\quad v_c =F_{cb} v_b \,,
\end{eqnarray}
where $ F_{ba}=-\left(M_b +iJ S_b+iJ R_{bc} F_{cb}\right)^{-1} (B_{ba}
+iJ R_{ba})$ and $F_{cb} =-\left(M_c +iJ S_c\right)^{-1} (B_{cb} +iJ
R_{cb})$. This recursive procedure is possible in principle for any
$N$, although, in general not very practical, given that the exact
solution can also be obtained by simply inverting one matrix, $M$, as
in Eq.~(\ref{eq:ThuJan10223943CET2013}). Anyhow, obtaining the exact
solution becomes exceedingly cumbersome for a large number of
cavities, $N \gg 1$. In the following we therefore concentrate on
finding an approximate solution.

\section{Approximated solution to second order in $J$}

In order to find an approximate semi-analytical expression for the
steady state of a cavity, $v_a$, we expand both the correlators in $v$
and the set of equations
(\ref{eq:SunMar10123437CET2013})--(\ref{eq:TueMar12191621CET2013}) in
powers of $J$ up to second order. More precisely, second order for
$v_a$, requires for consistency the following lower orders in the
other subsets:
\begin{eqnarray}
  \label{eq:ThuJan10222413CET2013}
  v_a&=& v_a^{(0)}+J v_a^{(1)}+J^2 v_a^{(2)}+\ldots \,,\\
  v_b&=&v_b^{(0)}+Jv_b^{(1)}+\ldots\,,\\
  v_c&=&v_c^{(0)}+\ldots\,,\\
  v_d&=&v_d^{(0)}+\ldots\,,\\
  v_e&=&v_e^{(0)}+\ldots\,.
\end{eqnarray}
The expanded equations read in these terms:
\begin{eqnarray}
  \label{eq:ThuJan10214848CET2013}
  \partial_t v_a=0 & =&  \Big[M_a v_a^{(0)}+I_a \Big] + J \Big[ M_a
  v_a^{(1)}+iS_a v_a^{(0)}+i R_{ab} v_b^{(0)} \Big] \nn \\
  &+&J^2 \Big[ M_a
  v_a^{(2)}+iS_a v_a^{(1)}
  +i R_{ab} v_b^{(1)} \Big]+\ldots\,,\\
  \partial_t v_b =0&=&\Big[ M_b v_b^{(0)}+B_{ba} v_a^{(0)} \Big] \nn \\
  &+& J \Big[ M_b v_b^{(1)}+iS_b v_b^{(0)} +B_{ba} v_a^{(1)} +i R_{ba}
  v_a^{(0)} +i R_{bc} v_c^{(0)}\Big]+\ldots \,.
\end{eqnarray}
We truncate the equations at this point since the
solutions obtained by setting each square bracket to zero,
\begin{eqnarray}
  \label{eq:ThuJan10222622CET2013}
  v_a^{(0)} &=&-M_a^{-1}I_a \,, \quad v_b^{(0)} =-M_b^{-1}B_{ba} v_a^{(0)}\,,\label{eq:ThuJan10222629CET2013}\\
  v_a^{(1)} &=&-M_a^{-1}\Big[ iS_a v_a^{(0)}+i R_{ab}
  v_b^{(0)}\Big] \,,\\
  v_b^{(1)} &=&-M_b^{-1}\Big[i R_{ba}
  v_a^{(0)} +B_{ba} v_a^{(1)}+ iS_b v_b^{(0)} +i R_{bc} v_c^{(0)}\Big]\,,\label{eq:ThuJan10222633CET2013}\\
  v_a^{(2)} &=&-M_a^{-1}\Big[iS_a v_a^{(1)} +i R_{ab}
  v_b^{(1)} \label{eq:ThuJan10222637CET2013} \Big]\,,
\end{eqnarray}
ultimately depend on $v_c^{(0)}$ only. Obtaining $v_c^{(0)}$ from the
equations would in turn require the knowledge of $v_e^{(0)}$ but this
is not needed given that the zero order is simply the uncoupled limit,
that is, products of the solutions for $N=1$  as in
Eq.~(\ref{eq:FriMar1184432CET2013}).  For instance, the uncorrelated
solution for $v_b^{(0)}$, corresponding to $\mean{a^{\dagger
    m}a^{n}b^{\dagger \mu}b^{\nu}}^{(0)}=\mean{a^{\dagger
    m}a^{n}}\mean{a^{\dagger \mu}a^{\nu}}$, can be directly obtained
through the product of twice $v_a^{(0)}$, as
$v_b^{(0)}=v_a^{(0)}X_{b}v_a^{(0)}$, where $X_b$ is the corresponding
mixing matrix obtained by inspection. This is completely equivalent to
the linear algebra solution of Eq.~(\ref{eq:ThuJan10222629CET2013}). The
same applies for $v_c^{(0)}$ but with two mixing matrices, $v_c^{(0)}
=v_a^{(0)}X_{c1}v_a^{(0)}X_{c2}v_a^{(0)}$.

These solutions are valid for $N\geq 4$, since adding more cavities to
the circle does not produce any structural qualitative change to
second order in $J$. The approximation is better the larger the
$N$. It is formally the same for $N=2$ and 3 (setting $v_c^{(0)}=0$
for $N=2$), but differs quantitatively to first and second order,
respectively due to different coefficients in the equations.

\section{Comparison between the exact and approximated results}

Since the approximated solutions are single valued, this method cannot
reproduce regimes where several steady states are compatible for the
individual system (corresponding to different steady states of the
ensemble) or any other instability regions like lasing. Its
perturbative nature allows it only to describe regimes where the
coupling is smaller than the effective decoherence or driving.  More
precisely, the \emph{weak coupling regime}, where new collective
eigen-modes are not required to describe the ensemble dynamics.

\begin{figure}[t] 
  \centering
\includegraphics[width=\linewidth]{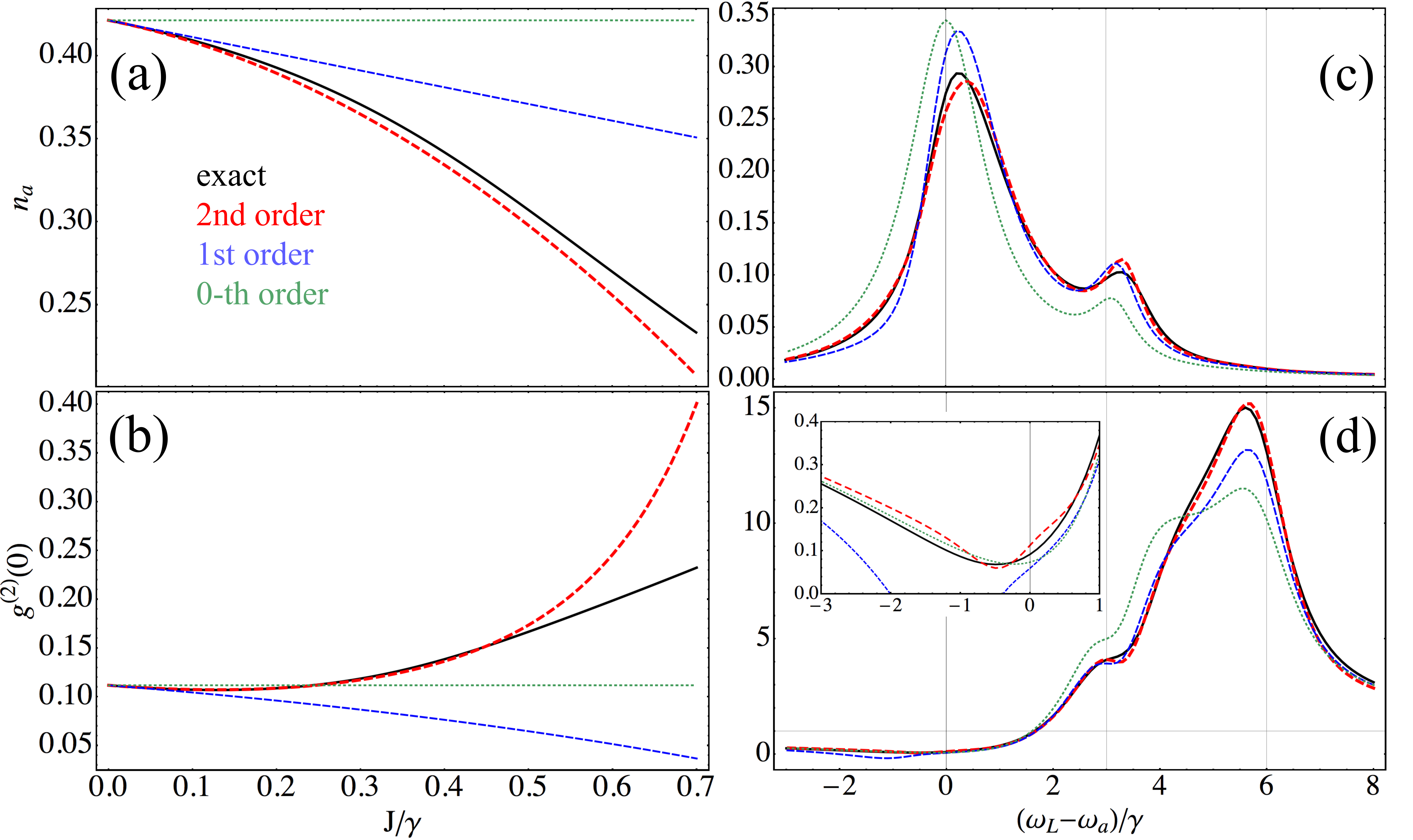}
\caption{Mean cavity population, $n_a$, and second order coherence
  function, $g^{(2)}(0)$. The exact solution for $N=4$ cavities (solid
  black) is compared to second (dashed red), first (dashed blue) and
  zero (dotted green) order approximations, valid for $N\geq 4$.  In
  (a) and (b) we fix $\Omega=0.7\gamma$ and $\Delta=0$ and vary the
  photon tunneling rate $J$. In (c) and (d) we fix $\Omega=0.5\gamma$
  and $J=0.3\gamma$ and vary the laser frequency
  $\omega_\mathrm{L}$. In the inset of (d), we have magnified the
  region around 0. Other parameters: $U=6\gamma$, $P=0$,
  $n_\mathrm{max}=2$.}
  \label{fig:1}
\end{figure}

We illustrate the interest of this method by comparing in
Fig.~\ref{fig:1} the exact solution for $N=4$ (in solid black) with
the approximated ones, valid for $N\geq 4$. We have chosen two
quantities of interest, the mean cavity population,
$n_a=\mean{a^\dagger a}$, and its second order coherence function at
zero delay, $g^{(2)}(0)=\mean{a^\dagger a^\dagger a
  a}/n_a^2$. Figs.~\ref{fig:1}(a) and (b) show that both are well
approximated by the second order solution (in dashed red) as long as
$J<\Omega,\gamma$. Lower order approximations (in blue and green)
deviate from the exact solution at even lower $J$. In
Figs.~\ref{fig:1}(c) and (d), we fix $J=0.3\gamma<\Omega=0.5\gamma$,
and scan the system resonances by tuning the laser frequency. In this
case, we observe that the second order approximation remains very
close to the exact solution for all frequencies while the first and
zeroth order deviate from it, notably, close to the various cavity
resonances at $0$, $U/2$, $U$ (marked with vertical lines). The first
order approximation breaks down near the one-photon resonance at $0$
as evidenced by the negative value of $g^{(2)}(0)$ in the zoom-inset
of Fig.~\ref{fig:1}(d).

\begin{figure}[t] 
  \centering
\includegraphics[width=0.5\linewidth]{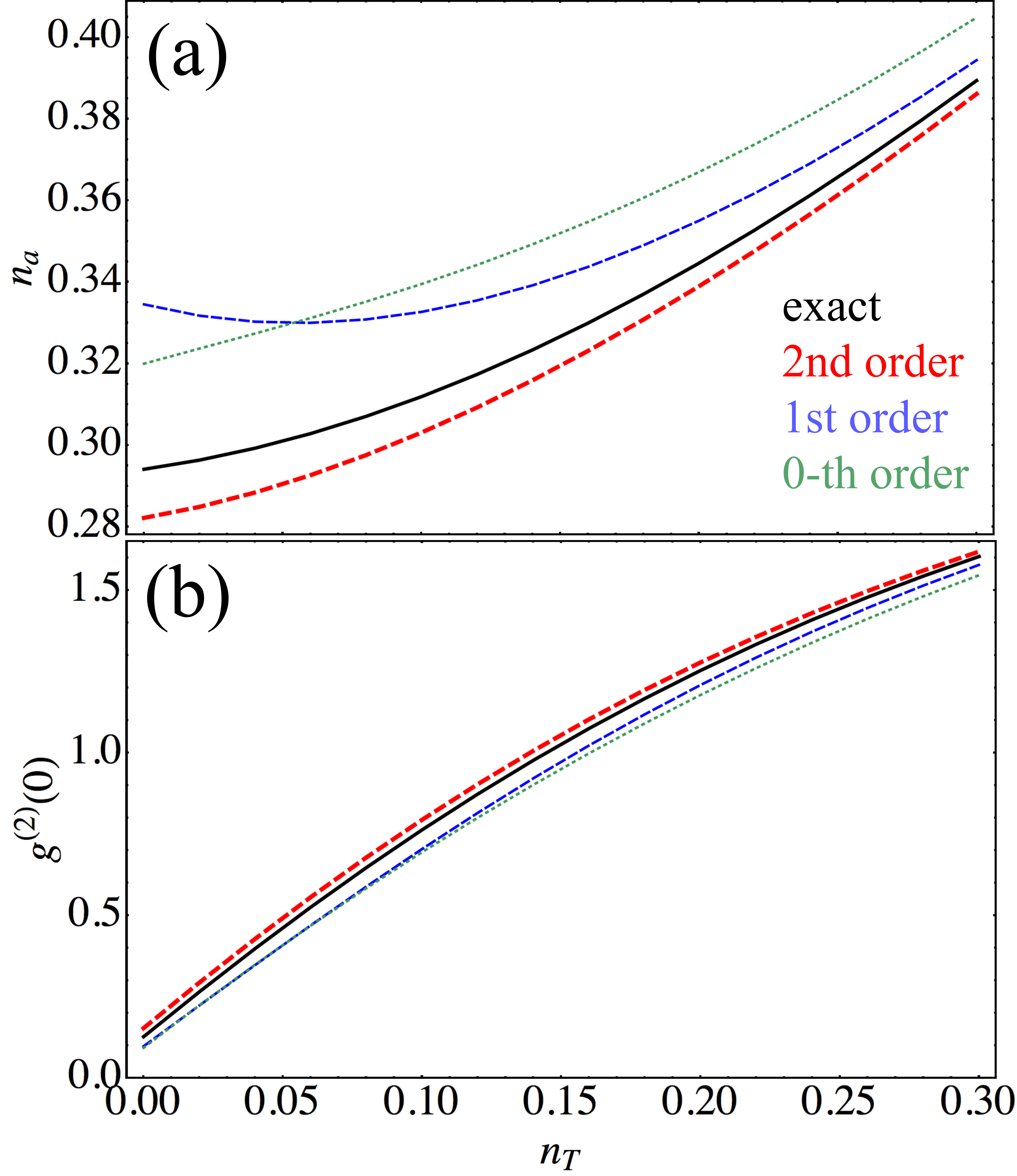}
\caption{(a) Mean cavity population, $n_a$, and (b) second order
  coherence function, $g^{(2)}(0)$, as a function of the thermal bath
  occupation. The exact solution for $N=4$ cavities (solid black) is
  compared to second (dashed red), first (dashed blue) and zero
  (dotted green) order approximations, valid for $N\geq 4$. Parameters
  are chosen to maximize the cavity population in Fig.~\ref{fig:1}(c):
  $U=6\gamma_0$, $\Omega=0.5\gamma_0$, $J=0.3\gamma_0$,
  $\omega_\mathrm{L}=\omega_a+0.25\gamma_0$ and $n_\mathrm{max}=2$.}
  \label{fig:2}
\end{figure}

Due to the form of the dissipation terms in
Eq.~(\ref{eq:ThuJan10202150CET2013}) we cannot illustrate the method
in the absence of a coherent drive, only under the action of an
incoherent pumping.  An incoherent pump bringing the system into a
steady state is equivalent to letting each cavity interact with a
thermal bath, where $P=n_\mathrm{T}\gamma_0$ and
$\gamma=(1+n_\mathrm{T})\gamma_0$. Here $\gamma_0$ is the decay rate
at zero temperature and $n_\mathrm{T}$ the occupation number of the
bath at temperature $T$, which are identical for all cavities in our
considerations. Hence for the case of purely dissipative dynamics with
$H_{N} = 0$ in Eq.~(\ref{eq:ThuJan10202150CET2013}), the steady state
$\tilde \rho_{\textrm{th}}$ is a product of thermal states at
temperature $T$ for each cavity, that is, $\tilde{\rho}_{\textrm{th}} = Z^{-1}
\sum_{n} e^{-\omega_{a} n / (k_{B} T)}$, where $k_{B}$ is Boltzmann
constant, $n$ the total number of excitations in the system and $Z$
the partition sum.  For $\Omega = 0$ the Hamiltonian in
Eq.~(\ref{eq:ThuJan10202154CET2013}) conserves the number of
excitations in the system and thus $[H_{N},\tilde{\rho}_{\textrm{th}}]
= 0$ so that $\tilde{\rho}_{\textrm{th}}$ is the steady state even in
the presence of unitary dynamics generated by $H_{N}$, independently
of the value of $J$.  Of course the rotating wave approximation that
has been applied to derive the Hamiltonian
(\ref{eq:ThuJan10202154CET2013}) is only valid for $U$, $J \ll
\omega_a$.  In this regime a thermal bath as described by the
dissipation terms in Eq.~(\ref{eq:ThuJan10202150CET2013}) is ``blind''
to the energy scales $U, J$ and all cavities will eventually be in
thermal equilibrium with their bath, $n_a=n_\mathrm{T}$ and
$g^{(2)}(0)=2$, regardless of the hopping $J$ and therefore other
neighboring cavities. The dynamics and spectrum of emission (out of
the scope of the present study) do, however, depend on the microscopic
properties of the cavities.  For example, larger $J$ and $\gamma_0$
would accelerate the thermalization of the cavity array.

If, on the other hand, a coherent and an incoherent drive are both
present, the steady state becomes nontrivial. Moreover, our approach
is suited for exploring this experimentally relevant scenario that
describes coherently driven cavities in the presence of thermal
background radiation.  Fig.~\ref{fig:2} shows $n_a$ and $g^{(2)}(0)$
as a function of the thermal bath occupation number $n_\mathrm{T}$,
which increases with increasing temperature, for the case of maximum
cavity population in Fig.~\ref{fig:1}(c). Both the exact and the
approximated solutions converge at high temperature to the thermalized
steady state ($n_a=n_\mathrm{T}$ and $g^{(2)}(0)=2$). At low
temperatures, where the emission is antibunched and each cavity
behaves like a single-photon emitter, first and zero order
approximations differ strongly from the exact solution, especially
when computing $n_a$, while, again, the second order approximation
follows it quite smoothly.

It is interesting to note that the steady state cavity spectrum of
emission could also be obtained from this approach without recurring
to the quantum regression theorem and, therefore, to deriving any time
dynamics. The alternative to such complications is to look into the
steady state occupation, as a function of its natural frequency, of a
mode, that is weakly coupled to the cavity array and which plays the
role of the detector. We showed with coworkers the equivalence between
this quantity and the power spectrum~\cite{delvalle12a}.

\section{Conclusions}

We have presented a method to solve for the steady state of coupled
cavities in a circular 1D array with translational symmetry, to second
order in the photon tunneling rate, $J$. This method can be
generalized to any set of identical weakly coupled systems, being in a
1D, 2D or 3D arrangement.

We consider any type of driving of the cavities (coherent or
incoherent), dissipation and nonlinearities. We first derive the
equations of motion for a minimal set of relevant correlators, $v_a$,
and then perform a power expansion of both the equations and the
solutions to obtain semi-analytical expressions for $v_a\approx
v_a^{(0)}+J v_a^{(1)}+J^2 v_a^{(2)}$. The approximated solution is
invariant for $N\geq 4$ cavities due to the nearest neighbor nature of
the coupling. We have finally illustrated the performance of our
method with an example of four weakly coupled cavities under a
coherent drive and temperature.

\section{Acknowledgements}

We acknowledge fruitful discussions with Peter Degenfeld-Schonburg. EdV
acknowledges support from the Alexander von Humboldt foundation and
MJH from the German Research Foundation (DFG) via the Emmy Noether
project HA 5593/1-1 and the CRC 631.

\section*{Appendix: Equations for the correlators}

In this appendix we provide the matrix $\tilde M$ and vector $\tilde
I$ appearing in Eq.~(\ref{eq:matrixform}), which are the starting
point of the described procedure. Let us consider operators for only
two adjacent cavities, $\mean{a_1^{\dagger m} a_1^n a_2^{\dagger \mu}
  a_2^\nu}$, as the general case of an array is an straight-forward
generalization. Then, we can obtain their equations of motion from the
full master equation as
\begin{equation}
  \label{eq:TueMay5174356GMT2009}
  \partial_t \mean{a_1^{\dagger m} a_1^n a_2^{\dagger \mu}
    a_2^\nu}=\Tr(\partial_t \rho \, a_1^{\dagger m} a_1^n a_2^{\dagger \mu}
  a_2^\nu) =\sum_{k,p,\alpha,\beta} \tilde R_{ \tiny
\begin{array}{c}
m,n,\mu,\nu\\
k,p,\alpha,\beta
\end{array}}\mean{a_1^{\dagger k} a_1^p a_2^{\dagger \alpha}
    a_2^\beta}\,.
\end{equation}
The corresponding elements in~$\tilde R$ are given
by~\cite{delvalle_book10a}:
\begin{eqnarray}
  \label{eq:TueDec23114907CET2008}
  \tilde R_{ \tiny
\begin{array}{c}
m,n,\mu,\nu\\
m,n,\mu,\nu 
\end{array}}&=&i\Delta
  (m-n)-\frac{\gamma}2(m+n)+i\frac{U}2 [m(m-1)-n(n-1)]\nonumber\\
  &+&i\Delta (\mu-\nu)-\frac{\gamma}2(\mu+\nu)+i\frac{U}2
  [\mu(\mu-1)-\nu(\nu-1)]\,,
\end{eqnarray}
\begin{eqnarray}
  \tilde R_{ \tiny
\begin{array}{c}
m,n,\mu,\nu\\
m-1,n-1,\mu,\nu 
\end{array}}=P m n \:
 & , & \quad  \tilde R_{ \tiny
\begin{array}{c}
m,n,\mu,\nu\\
m,n,\mu-1,\nu-1 
\end{array}}=P \mu\nu\,,\\
  \tilde R_{ \tiny
\begin{array}{c}
m,n,\mu,\nu\\m+1,n+1,\mu,\nu 
\end{array}}=iU( m-n) \:
 & , & \quad  \tilde R_{ \tiny
\begin{array}{c}m,n,\mu,\nu\\m,n,\mu+1,\nu+1 
\end{array}}=iU (\mu-\nu)\,,\\
  \tilde R_{ \tiny
\begin{array}{c}
m,n,\mu,\nu\\
m-1,n,\mu,\nu 
\end{array}}=i\Omega m \:
 & , & \quad  \tilde
  R_{ \tiny
\begin{array}{c}
m,n,\mu,\nu\\
m,n,\mu-1,\nu 
\end{array}}=i\Omega \mu \,,\\
  \tilde R_{ \tiny
\begin{array}{c}
m,n,\mu,\nu\\
m,n-1,\mu,\nu 
\end{array}}=-i\Omega n \:
 & , & \quad  \tilde
  R_{ \tiny
\begin{array}{c}
m,n,\mu,\nu\\
m,n,\mu,\nu-1 
\end{array}}=-i\Omega \nu \,,\\
  \tilde R_{ \tiny
\begin{array}{c}
m,n,\mu,\nu\\
m-1,n,\mu+1,\nu 
\end{array}}=iJ m \:
 & , & \quad  \tilde
  R_{ \tiny
\begin{array}{c}
m,n,\mu,\nu\\
m+1,n,\mu-1,\nu 
\end{array}}=i J  \mu \,,\\
  \tilde R_{ \tiny
\begin{array}{c}
m,n,\mu,\nu\\
m,n-1,\mu,\nu+1 
\end{array}}=-iJ n \:
 & , & \quad  \tilde
  R_{ \tiny
\begin{array}{c}
m,n,\mu,\nu\\
m,n+1,\mu,\nu-1 
\end{array}}=-i J  \nu \,,
\end{eqnarray}
and zero everywhere else. The vector $\tilde I$ is constructed from
the elements that provide an independent term for the equations, that
is, $\tilde R_{ \tiny
\begin{array}{c}
m,n,\mu,\nu\\
0,0,0,0 
\end{array}}$, while the matrix
$\tilde M$ corresponds to all other elements. 

\section*{References}

\end{document}